\documentclass[aps,prb,twocolumn,superscriptaddress]{revtex4}
\usepackage{ae}
\usepackage[T1]{fontenc}
\usepackage[ansinew]{inputenc}
\usepackage{amsmath}
\usepackage{amssymb}
\usepackage{graphicx}
\usepackage{color}
\usepackage[colorlinks]{hyperref}
\usepackage{epstopdf}

\begin{document}

\date{\today}

\title{Onset of pseudogap and density wave in a system with a closed Fermi surface}

\author{M. Spai\'{c}}
\author{D. Radi\'{c}*}
\affiliation{Department of Physics, Faculty of Science, University of Zagreb, Bijeni\v{c}ka 32, Zagreb 10000, Croatia}
\date{\today}

\begin{abstract}
We study the influence of anisotropy, treated as a dimensional crossover between 1D and 3D system, on the topological instability induced by a (self-consistent) uniaxial periodic potential. The mechanism on which the instability is based involves the topological reconstruction of the Fermi surface, from initially closed pockets to the surface with open Fermi sheets, creating two peculiar points in the band dispersion - the saddle point and elliptical point, between which the pseudogap in electron density of states develops. The self-consistent periodic potential appears as a result of interactions, either electron-phonon, or electron electron, which, linked with the topological instability of the system, results in formation of a new ground state of the system - the density wave provided that the relevant coupling constant is larger than critical.
Our analysis shows that the phase transition takes place along the whole continuous interval of a dimensional crossover between 1D and 3D, but that the critical coupling strength significantly increases with the dimensionality of the system.
It is our intention to give an initial framework for understanding the nature of charge density waves experimentally observed in a number of materials, like high$-T_c$ cuprates or graphite intercalates, both being materials with a closed, rather isotropic Fermi surface far from the nesting regime.
\end{abstract}

\maketitle

\bigskip
\textbf{I. Introduction}
\bigskip

The density wave (DW), either of electron charge (CDW) or spin (SDW), is a periodic modulation of a crystal that appears spontaneously as a result of structural instability. First experimental observations of DWs were in low-dimensional organic conductors from the family of Bechgaard salts, which later extended to the wide class of materials with a highly anisotropic Fermi surface (FS) \cite{Gruner, Pouget, Thorne}. The onset of these phenomena lays in Peierls instability by which, due to opening of a gap at the Fermi surface, the metallic ground state of a one-dimensional (1D) conductor becomes unstable with respect to the formation of a self-consistent periodic modulation of the electron charge - CDW that constitutes a potential with a wave vector exactly relating two points of the Fermi surface in 1D \cite{Peierls}. The possibility to completely gap the FS exists not only in 1D systems, but also in two-dimensional (2D) ones with high degree of anisotropy, called ``quasi-one-dimensional" (Q1D), such that parts of the 2D open FS can be completely mapped to each other by a unique wave vector. The mechanism based on such property of the FS is called ``nesting" and it has been the raw model to explain the DW physics for last several decades. Even in the absence of perfect nesting, i.e., when mapping of the Fermi sheets leaves pockets of states, small enough as compared to squared inverse magnetic field length, one can utilize external magnetic field to ``fix" the nesting through the Landau quantization mechanism, leading to field-induced DW phases \cite{FISDW}. Due to the unique nature of the Peierls mechanism to completely gap the FS, the condensation energy of the DW contains the characteristic logarithmic contribution which, in turn, gives at zero temperature the finite order parameter even for arbitrarily small coupling responsible for the DW formation (electron-phonon, electron-electron, etc.) \cite{Solyom}.\\

The natural question that arises is whether it is possible to stabilize the DW ground state unless the FS has the property of nesting? The experimental findings are very affirmative in that respect. The CDWs, often called ``charge stripes", are detected in number of 2D and quasi-2D compounds with rather isotropic FSs far away from the nesting possibility, among which we single out the high-temperature superconducting cuprates, with conducting CuO$_2$ planes \cite{cuprates}, graphite-based intercalates like CaC$_6$ \cite{graphene}, and transition metal dichalcogenides (TMDs) like 2H-NbSe$_2$ \cite{TMD} as the most intriguing examples. Despite numerous experimental findings appearing for more than a decade, the feasible theoretical framework to explain such DWs, outside of the nesting physics, has not been established. For example, experimental observations in 2H-NbSe$_2$ \cite{Tonjes}, exhibiting both nested- and saddle-band contributions on the FSs, fits partially to the nesting mechanism (wave vector of the CDW) and partly to the ``saddle point mechanism" \cite{RiceScott} (resistivity decrease in the CDW transition), but none of them providing  understanding of the complete picture.\\

Starting from the 1D system, there are two possible scenarios through which the electron band can lower the energy of the condensate by imposing the self-consistent DW potential upon itself. The first one we call ``the Peierls scenario", in which the system attempts to gap the whole Fermi surface through the nesting mechanism as described above. The second one we call ``the Lifshitz scenario" in which system changes the topology of the FS, manipulating electron density of states (DOS) to push more populated states to lower energies compared to the situation before the topological change. Recently, it has been predicted that topological reconstruction of the FS in 2DEG, from the isotropic closed contour to the open one, may lead to lowering of total energy of electron band and consequently to the quantum phase transition into the DW ground state if the relevant coupling constant (electron-phonon in the presented case) is larger than critical \cite{KBR}. This DW state persisted and was even strengthened in homogeneous external magnetic field \cite{KBR2}, stabilizing with the wave vector $Q \approx 2k_F(1-\Delta/2\epsilon_F)$, where $k_F$ is the Fermi wave vector, $\epsilon_F$ the Fermi energy and $\Delta$ is the gap in energy spectrum. In the cited works, the DW order parameter $\Delta$ is treated as a gap parameter, although the FS is not actually fully gapped. Topological reconstruction of the initially parabolic 2DEG band spectrum, leads to the emergence of two peculiar points in the final spectrum, below and above the initial Fermi energy: hyperbolic point in the lower and elliptical point in the upper new band. They appear at the same wave vector, one upon another, separated in energy by $2\Delta$. Due to similarity of the presented band reconstruction mechanism with the mechanism behind Lifshitz's topological transition ``of $2 \frac{1}{2}$ order" \cite{Lifshitz, Volovik}, we call these peculiar points the ``Lifshitz" points. Along the energy interval between ``Lifshitz" points the pseudogap in the density of states develops. Due to the structure of the pseudogap, in particular, the van Hove singularity of logarithmic type around the hyperbolic point, redistribution of states to lower energies decreases the total band energy. \\

In this work, we generalize the picture of topological reconstruction of electron band in dimensional (noninteger) crossover starting from 1D to 3D system, and find the conditions necessary to establish the DW ground state. In Section II, we present the model; the change in electron DOS accompanying the reconstruction process  and being the precursor of dramatic change of the ground state is presented in section III; the results for condensation energy of the new ground state and optimal values for its Fermi energy, wave vector, order parameter and critical coupling constant are presented in Section IV; the discussion and final conclusions are in Section V.

\bigskip
\textbf{II. Model}
\bigskip

We consider the system generally described by a form of Hamiltonian $H=H_0+H_{int}$, where $H_0$ describes the one-particle contributions (e.g., electron and phonon bands), while $H_{int}$ describes the interacting electrons (e.g. electron-phonon, or electron-electron interaction). We assume an existence of a self-consistent, periodic order parameter, manifesting itself in the form of uniaxial density wave with wave vector $\textbf{Q}$. Such an order parameter usually has an origin in a nonvanishing expectation value of either phonon displacement operator, or electron density operator due to macroscopically large population of the corresponding state characterized by the wave vector $\textbf{Q}$. However, regardless of the nature of that order parameter the effective Hamiltonian of the system, within the mean-field approximation and some simplifications specific for particular type of interaction \cite{Solyom}, can be written in the form
%
\begin{align}
H=&\sum_{\textbf{k}} {\left[ \epsilon(\textbf{k}) a_{\textbf{k}}^{\dag} a_{\textbf{k}} + \Delta e^{i\phi} a_{\textbf{k}+\textbf{Q}}^{\dag} a_{\textbf{k}} + \Delta e^{-i\phi} a_{\textbf{k}-\textbf{Q}}^{\dag} a_{\textbf{k}}\right]} \nonumber \\
 &+ \frac{\Delta^2}{G}.
\label{MFHamiltonian}
\end{align}
%
Here, $a_{\textbf{k}}$ and $a_{\textbf{k}}^{\dag}$ are electron annihilation and creation operator in the state with wave vector $\textbf{k}$, $\epsilon(\textbf{k})$ is the one-electron band dispersion, while $\Delta$ and $\phi$ are the amplitude and phase of the order parameter. Let us choose the direction of periodic uniaxial DW perturbation along the $x-$axis, i.e. $\textbf{Q}=(Q,0,0)$, such that the potential created by the perturbation is $\Delta \cos(Qx+\phi)$. In the system without pinning the phase $\phi$ is arbitrary. Finally, $G$ is the coupling constant which parametrizes electron coupling entering the problem, e.g., Fr\"{o}hlich, for electron-phonon, or Hubbard, for electron-electron coupled systems \cite{Solyom,KBR}. The dependence of the coupling constant on $\textbf{Q}$ is assumed to be slow enough to be considered constant and neglected in condensation energy optimization. The dependence of the order parameter on $\textbf{Q}$ is assumed implicitly, adopting the label $\Delta_{\textbf{Q}} \rightarrow \Delta$ for the sake of presentation, and will be treated later.\

The electronic part (first term) of Hamiltonian (\ref{MFHamiltonian}) can be diagonalized in the straightforward way \cite{Solyom} providing the spectrum
%
\begin{align}
E_{\pm}(\mathbf{k})= \frac{\epsilon(\mathbf{k})+\epsilon(\mathbf{k+Q})}{2}\pm
 \sqrt{\Big(\frac{\epsilon(\mathbf{k})-\epsilon(\mathbf{k+Q})}{2} \Big)^2+\Delta^2}, \nonumber \\
\label{NewBands}
\end{align}
%
for electron operators $\tilde{a}_{\textbf{k}\alpha}$ and $\tilde{a}_{\textbf{k}\alpha}^{\dag}$, reducing the Hamiltonian to the form
%
\begin{align}
H=\sum_{\textbf{k},\alpha=\pm}{E_{\alpha}(\textbf{k}) \tilde{a}_{\textbf{k}\alpha}^{\dag} \tilde{a}_{\textbf{k} \alpha}} + \frac{\Delta^2}{G}.
\label{MFHamiltonian2}
\end{align}
%
We see that the order parameter $\Delta$ appears as a gap parameter in one-electron spectrum (\ref{NewBands}). Using, for the sake of simplicity, the free-electron dispersion, with effective mass $m$, as $\epsilon(\textbf{k})$ in further description, and shifting the origin of momentum space by $- Q/2$ along the $k_x$ axis (i.e., placing it at the edge of the new Brillouin zone), the $\mathbf{Q}-$related $\epsilon(\textbf{k})$ terms in Eq. (\ref{NewBands}) read
%
\begin{align}
\epsilon(k_x \pm Q/2, k_y, k_z)
=\frac{\hbar^2 \left(k_x\pm Q/2\right)^2}{2m}+\frac{\hbar^2 k^2_y}{2m}+\frac{\hbar^2 k^2_z}{2m},
\label{ShifDispersion}
\end{align}
%
while the diagonalized electron spectrum, Eq.(\ref{NewBands}), attains the form
%
\begin{align}
E_{\pm}(\mathbf{k})= \frac{\hbar^2}{2m} & \left( \frac{Q^2}{4} +k_x^2 + k_y^2+k_z^2 \right. \nonumber \\
&\left. \pm \sqrt{(Qk_x)^2+\left( \frac{2m \Delta}{\hbar^2} \right)^2} \right). \nonumber \\
\label{NewBandsSpectrum}
\end{align}
%

\begin{figure}
\centerline{\includegraphics[width=\columnwidth]{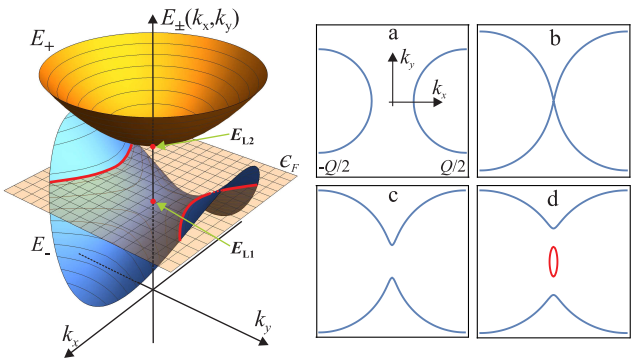}}
\caption{An illustration of topological reconstruction of the Fermi surface, for simplicity of graphical presentation shown for the case of 2D parabolic band. The left panel shows the spectrum $E_{\pm}(k_x,k_y)$ of the reconstructed band (\ref{NewBandsSpectrum}) around the peculiar points L$_1$ and L$_2$, where $\epsilon_F$ is the Fermi energy (plane). The cross section $E_{\pm}(k_x,k_y)=\epsilon_F$ determines the Fermi surface (thick curves). The right panel shows the Fermi surface in the process of reconstruction depending on the Fermi energy (blue for $E_{-}$, red for $E_{+}$): (a) $\epsilon_F<E_{L1}$, (b) $\epsilon_F=E_{L1}$, (c) $E_{L1}<\epsilon_F<E_{L2}$ and (d) $\epsilon_F>E_{L2}$.}
\label{Figbands}
\end{figure}

Spectrum (\ref{NewBandsSpectrum}) contains two peculiar points (``Lifshitz points") at the origin $\mathbf{k}=0$: L$_1$ - the saddle (hyperbolic point) in the lower band at energy $E_{L1}=\hbar^2Q^2/8m-\Delta$, and L$_2$ -  parabolic minimum (elliptic point) at energy $E_{L2}=\hbar^2Q^2/8m+\Delta$ (see Fig. \ref{Figbands} for illustration in the 2D case).

%
%

\bigskip
\textbf{III. Electron density of states (DOS)}
\bigskip

In order to track the influence of the band reconstruction on the properties of the system the most basic and useful quantity to find is electronic density of states (DOS). Furthermore, to grasp the aspect of anisotropy of the (initial) system, we introduce the effective dimensionality. The concept of effective dimensionality, namely, in the momentum space, is to address the degree of confinement of electron motion in certain dimension(s). Such effective dimension does not necessarily need to be an integer number. It is known from the early concepts of discrete lattices with non-integer effective dimension \cite{Dhar}, proteins \cite{Stapleton, Elber}, interaction-generated synthetic dimensions in cold atoms physics \cite{Lewenstein} and optics \cite{Boada,Buljan}, to nowadays very active field of complex networks \cite{Daqing,Shanker}.
One possible approach to address anisotropy in the solid state physics is to utilize the tight binding approximation (TBA) with different electron hopping integrals along different spatial directions, showing to reproduce electron DOS interpolating between 1D and 3D \cite{Nakao, Okuyama}.
Another possible approach, suggested by He \cite{He1990}, is based on the heuristic idea that one can describe motion, with anisotropy in the sense of certain restrictions in some dimension(s) of the 3-dimensional Euclidean embedding space, as the isotropic motion in the deformed, effectively $d-$dimensional space, where $d\le 3$ may be non-integer, reflecting the reduction in number of accessible states for such motion. This concept has been used in literature to describe various physical effects in systems with different types of confinement of electron motion, e.g. impurities in semiconductor superlattices \cite{Gomez}, optical transitions in anisotropic solid \cite{He1991}, or exciton-phonon interaction \cite{Thilagam}, free exciton binding \cite{Zhao, Mathieu} and polaronic effect \cite{Matos} in the quantum wells, etc. The calculated electron DOS is in a good agreement with the afore-mentioned one calculated by the TBA method and, also, represents a simple phenomenological tool to fit experimental observations where anisotropy is manifested, e.g., the measured heat capacity of anisotropic 3D lattice $C_v \sim T^d$ at low temperature $T$, exhibiting the fractional values of $d$ between 2 and 3 within the crossover temperature range as the 2D confinement deteriorates and the system becomes 3D \cite{Keesom}. We have to emphasize that such approach goes beyond the mere effective mass description, which changes mass dependence in DOS, i.e., $m^D \rightarrow \prod_{i=1}^D m_i^*$ ($D$ is standard integer dimensionality), but it does not change the dependence on energy as compared to the corresponding isotropic system \cite{Solyom-DOS}.\\

Here we introduce the dimensionality parameter $d$ that can change continuously within the interval between 1 and 3, where values $d=1$, 2 and 3 correspond to the strictly 1D, 2D and 3D standard dimensionality, while the values between 1 and 2 account for so to say quasi-1D, and between 2 and 3 for quasi-2D systems.
The electron DOS in our problem is calculated as
%
\begin{align}
g(E)=\frac{2}{(2 \pi)^{d}} \frac{{\mathrm d} \Omega_k^{(d)}(E)}{{\mathrm d}E}, 
\label{DOS-def2}
\end{align} 
%
where $\Omega_k^{(d)}(E)$ is a volume of $d-$dimensional reciprocal space enclosed within the surface of constant energy $E$, 2 in numerator accounts for the spin degeneracy. We need to stress here that electron DOS (\ref{DOS-def2}) depends, beside on energy $E$, also on number of other parameters, namely, $d$, $\Delta$, $Q$. For the purpose of presentation, we assume them implicitly, i.e., adopt the label $g^{(d)}_{\Delta,\mathbf{Q}}(E) \rightarrow g(E)$ in the text, but showing and emphasizing the corresponding dependence as it appears in the text. In the unperturbed system with no band reconstruction ($\Delta=0$), volume $\Omega_k^{(d)}(E)$ reduces to
%
\begin{align}
\overline{\Omega}_k^{(d)}(E)=\frac{\pi^{d/2}}{\Gamma \left( \frac{d}{2}+1 \right)} E^{\frac{d}{2}}, 
\label{Omega0}
\end{align} 
%
where $\Gamma(z)$ is the standard gamma function. Eq. (\ref{Omega0}) is obtained from the volume of the $d-$dimensional sphere with radius $k$ in the reciprocal space, $\overline{\Omega}_k^{(d)}=\frac{\pi^{d/2}}{\Gamma \left( d/2+1 \right)} k^{d}$, using the free-electron dispersion $E(k)= \frac{\hbar^2 k^2}{2m}$.  Using Eqs. (\ref{DOS-def2}) and (\ref{Omega0}), it is straightforward to obtain standard expression for the DOS of the unperturbed band, i.e. of the free $d-$dimensional electron gas 
%
\begin{align}
g_0(E)=\frac{2m^{d/2}}{(2 \pi)^{d/2}\hbar^d \Gamma \left( \frac{d}{2} \right)} E^{\frac{d}{2}-1}, 
\label{DOS-g0}
\end{align} 
%
which exactly reproduces known cases with integer dimensionality \cite{123DEG}.\\

\begin{figure}
\centerline{\includegraphics[width=\columnwidth]{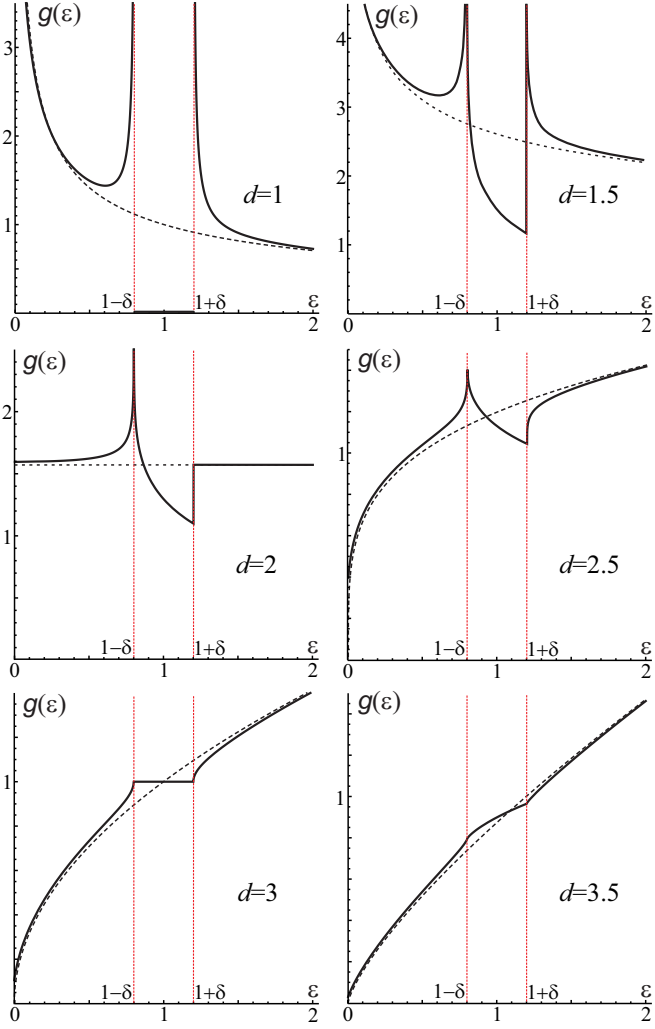}}
\caption{Electron density of states $g(\varepsilon)$, Eq.(\ref{DOS}), scaled by prefactor in front of the multiplication operator $\times$, depending on dimensionality parameter $d$. Full curve is the DOS of the reconstructed system with gap $\delta=0.2$, dashed is the DOS of the system without reconstruction ($\delta=0$), i.e. $g_0(\varepsilon)$ based on Eq. (\ref{DOS-g0}). Peculiar "Lifshitz" points L$_1$ and L$_2$ are at $1-\delta$ and $1+\delta$ respectively. Between them the pseudogap is formed (except for $d=1$ where the gap is real, i.e. $g(\varepsilon)=0$). The last panel shows the case for $d=3.5$, although outside of the physical range of $d-$values for this particular problem, but presented for the sake of depicting the analytical behaviour of the function $g(\varepsilon)$ for $d>3$.}
\label{FigDOS}
\end{figure}

We adopt the suitable scaling of variables to the dimensionless form, namely,
%
\begin{align}
\varepsilon \equiv \frac{2 m E}{\hbar^2(Q / 2)^{2}}, \quad \delta \equiv \frac{2 m \Delta}{\hbar^2 (Q / 2)^{2}}, \quad
\left( \begin{array}{c}
\kappa_x \\
\kappa_y \\
\kappa_z
\end{array}
\right) \equiv \frac{2}{Q}
\left( \begin{array}{c}
k_x \\
k_y \\
k_z
\end{array}
\right),
\label{variables-scaling}
\end{align}
%
where $\varepsilon$, $\delta$ and $(\kappa_x,\kappa_y,\kappa_z)$ stand for energy, gap parameter and wave vector components respectively. 
Now the electron dispersion (\ref{NewBandsSpectrum}) reads
%
\begin{align}
\varepsilon_{\pm}(\kappa_x,\kappa_y,\kappa_z)=1+\kappa_x^2+\kappa_y^2+\kappa_z^2 \pm \sqrt{4\kappa_x^2+\delta^2}.
\label{spectrum}
\end{align}
%
The volume in reciprocal space gets scaled as $\Omega_k^{(d)}(E)=(Q/2)^d \Omega_{\kappa}^{(d)}(\varepsilon)$. This volume possesses rotational symmetry around the $\kappa_x-$axis thus, choosing $\kappa_{\rho}^2\equiv \kappa_y^2+\kappa_z^2$, which using Eq.(\ref{spectrum}) can be expressed as
%
\begin{align}
\kappa_{\rho}^{\pm}(\kappa_x;\varepsilon)=\sqrt{\varepsilon-1-\kappa_x^2 \mp \sqrt{4\kappa_x^2+\delta^2}},
\label{rho}
\end{align}
%
we can calculate it as
%
\begin{align}
\Omega_{\kappa}^{(d)}(\varepsilon)=2\frac{\pi^{\frac{d-1}{2}}}{\Gamma \left( \frac{d-1}{2}+1 \right)}\left( \int_{\kappa_0^{-}(\varepsilon)}^1 \kappa_{\rho}^{-}(\kappa_x;\varepsilon)^{d-1} {\mathrm d}\kappa_x \right. \nonumber \\
\left. + \int_{0}^{\kappa_0^{+}(\varepsilon)} \kappa_{\rho}^{+} (\kappa_x;\varepsilon)^{d-1} {\mathrm d}\kappa_x \right),
\label{volume}
\end{align}
%
utilizing the fact that, with the given cylindrical symmetry of our problem, the function under integral is in fact the volume of $(d-1)-$dimensional sphere with radius $\kappa_{\rho}^{\pm}$.
Here
%
\begin{align}
\kappa_0^{-}(\varepsilon)=\left\lbrace  
\begin{array}{c}
\sqrt{\varepsilon +1 - \sqrt{4\varepsilon + \delta^2}}, \,\,\, \varepsilon <1-\delta \\
0, \quad \quad \quad \quad \quad \quad \quad \quad \varepsilon \ge 1-\delta
\end{array}
\right.
\label{xcM}
\end{align}
%
and
%
\begin{align}
\kappa_0^{+}(\varepsilon)=\left\lbrace  
\begin{array}{c}
0, \quad \quad \quad \quad \quad \quad \quad \quad  \varepsilon < 1+\delta \\
\sqrt{\varepsilon +1 - \sqrt{4\varepsilon + \delta^2}}, \,\,\, \varepsilon \ge 1+\delta
\end{array}
\right.
\label{xcP}
\end{align}
%
are integration limits due to the gap opening. Inserting the above into expression (\ref{DOS-def2}), carefully taking the derivative \cite{derivative} over scaled energy $\varepsilon$, we get the expressions for the DOS:\\
\begin{subequations}
(a) for $d=1$
%
\begin{align} 
g(\varepsilon)=\frac{2 m}{\pi \hbar^2} \left(\frac{Q}{2}\right)^{-1} \times
\left\lbrace 
\begin{array}{c}
-\frac{1-2/\sqrt{4\varepsilon + \delta^2}}{\sqrt{1+\varepsilon-\sqrt{4\varepsilon+\delta^2}}}, \,\,\, \varepsilon<1-\delta \\
0, \quad \quad \quad \quad \varepsilon \in (1-\delta,1+\delta) \\
+\frac{1-2/\sqrt{4\varepsilon + \delta^2}}{\sqrt{1+\varepsilon-\sqrt{4\varepsilon+\delta^2}}}, \,\,\, \varepsilon>1+\delta 
\end{array}
\right. \nonumber \\
\label{DOSd1}
\end{align}
%
(b) for $d>1$
%
\begin{align}
&g(\varepsilon)=\frac{8 m \left(\frac{Q}{2}\right)^{d-2}}{(2 \pi)^{d} \hbar^2} \frac{\pi^{\frac{d-1}{2}}}{\Gamma\left(\frac{d-1}{2}\right)} \times \nonumber \\
&\left( \int_{\kappa_0^{-}(\varepsilon)}^1 \kappa_{\rho}^{-} (\kappa_x;\varepsilon)^{d-3} {\mathrm d} \kappa_x +  \int_{0}^{\kappa_0^{+}(\varepsilon)} \kappa_{\rho}^{+} (\kappa_x;\varepsilon)^{d-3} {\mathrm d} \kappa_x \right). 
\label{DOSdgt1}
\end{align}
%
\label{DOS}
\end{subequations}
We present the electron DOS (\ref{DOS}) for various dimensionalities of the system $d$ in Fig. \ref{FigDOS}. As it is seen from the picture, only in the strictly 1D ($d=1$) regime the system exhibits a real gap in the energy interval between $1-\delta$ and $1+\delta$ (i.e. between $E_{L1}$ and $E_{L2}$). For $d>1$, the system exhibits a pseudogap within which the number of states between $E_{L1}$ and $E_{L2}$ is more depleted, the lower the dimensionality $d$ is. As we show in the forthcoming sections, it has the crucial role in stabilization of the DW ground state. The $d=1$ expression is given separately from the general $d>1$ expression since it represents a nonanalytical limit.\\

As for the singularities appearing in the peculiar points, for integer values of $d$, i.e. for 1D, 2D and 3D systems, it is easy to characterize them analytically. For $d=1$ case, the simple Taylor expansion around the L$_1$ point yields $g(\varepsilon) \sim |\varepsilon - \varepsilon_{L1}|^{-1/2}$ dependence. For $d=2$ case, the Taylor expansion of the function under the square root in denominator of the integrand around $\kappa_0^{-}$ gives zero of the first order for $\varepsilon < \varepsilon_{L1}$, while its vanishing first derivative at $\varepsilon = \varepsilon_{L1}$ makes the zero of the second order at that point. This implies singular contribution to the integral in the form $\int (\kappa-\kappa_{L1})^{-1}\mathrm{d}\kappa$, yielding the density of states with logarithmic singularity, i.e. $g(\varepsilon) \sim \ln |1-\varepsilon/ \varepsilon_{L1}|$. Finally, the $d=3$ case can be easily integrated exactly, i.e. $g(\varepsilon) \sim 1-\kappa_0^{-}+\kappa_0^{+}$, which yields $g(\varepsilon) \sim 1 \mp const \cdot \sqrt{|\varepsilon - \varepsilon_{L1,2}|}$, for $\varepsilon \lessgtr \varepsilon_{L1,2}$, and $g(\varepsilon) \sim 1$, for $\varepsilon \in (\varepsilon_{L1},\varepsilon_{L1})$. Such singularities are in accordance with Van Hove's general classification (around the saddle point for the particular dimensionality of the system) \cite{vanHove}.

%
%

\bigskip
\textbf{IV. Self-consistent optimal Fermi energy, wave vector and order parameter of the DW}
\bigskip

In order to find conditions for stabilization of the DW ground state at zero temperature, one needs to compare energy of the DW-ordered state of the system with reconstructed Fermi surface, $E_{rec}$, determined by Hamiltonian (\ref{MFHamiltonian2}), with the energy of the state without the DW ordering (here the free-electron ground state), $E_{0}$, and self-consistently determine the position of the new Fermi energy of the system $\epsilon_F$ with respect to the initial one $\epsilon_{F0}$, the DW wave vector $Q$, and the DW order parameter $\Delta$ by imposing the criterion of maximal energy decrease in the system reconstruction process. The difference of these two energies we define as the DW condensation energy, i.e. $E_{DW}(\epsilon_F,\epsilon_{F0},Q,\Delta)=E_{rec}(\epsilon_F,Q,\Delta)-E_{0}(\epsilon_{F0})$. The $E_{rec}$ is a sum of two contributions: the electron energy of reconstructed band $E_{band}$ and "elastic" energy $E_{elast}=\Delta^2/G$ that arises within the mean-field description as a (positive) energy increase due to the finite order parameter formation (e.g. elastic energy of the deformed crystal lattice in the CDW case). Taking the contributions described above into account, we can finally express the DW condensation energy as the sum of the change of electron band energy due to reconstruction, $\Delta E_{band}$, and the (always positive) ``elastic" energy, i.e., 
%
\begin{align}
E_{DW}(\epsilon_F,\epsilon_{F0},Q,\Delta)=\Delta E_{band}(\epsilon_F,\epsilon_{F0},Q,\Delta)+\frac{\Delta^2}{G},
\label{EDW-def}
\end{align}
%
where
%
\begin{align}
\Delta E_{band}(\epsilon_F,\epsilon_{F0},Q,\Delta)=E_{band}(\epsilon_F,Q,\Delta)-E_{0}(\epsilon_{F0}).
\label{DeltaEband-def}
\end{align}
%
The first step is finding the relation between $\epsilon_F$ and $\epsilon_{F0}$. Since these are contained only in $\Delta E_{band}(\epsilon_F,\epsilon_{F0},Q,\Delta)$, it is enough to minimize only that function. In that respect we use scaling (\ref{variables-scaling}) and write the change of the band energy in scaled variables as
%
\begin{align}
\Delta \mathcal{E}_{band} (\varepsilon_F,\varepsilon_{F0};\delta)=\int_0^{\varepsilon_F}{g(\varepsilon;\delta)\varepsilon \mathrm{d}\varepsilon}-\int_0^{\varepsilon_{F0}}{g_0(\varepsilon)\varepsilon \mathrm{d}\varepsilon},
\label{DeltaEband}
\end{align}
%
where we should have in mind implicit dependence on $Q$, since all variables are scaled with it. Here $g(\varepsilon;\delta)$ and $g_0(\varepsilon)$ is electron DOS for reconstructed and non-reconstructed band respectively.\\
We seek the minimum of this function under the constraint of conservation of number of electrons $N$, meaning that change of electron number in the reconstruction process must be zero, i.e. $\Delta N=0$, where
%
\begin{align}
\Delta N (\varepsilon_F,\varepsilon_{F0};\delta)=\int_0^{\varepsilon_F}{g(\varepsilon;\delta) \mathrm{d}\varepsilon}-\int_0^{\varepsilon_{F0}}{g_0(\varepsilon) \mathrm{d}\varepsilon}.
\label{DeltaN-def}
\end{align}
%
Here $\delta$ is kept as a parameter (not written explicitly further on for the sake of simplicity). In order to find the minimum of energy under the given constraint, we introduce the Lagrange multiplier $\mu$ and define the Lagrange function $\mathcal{L}=\Delta \mathcal{E}_{band} - \mu \Delta N$. Optimal relation of Fermi energies $\varepsilon_F$ and $\varepsilon_{F0}$ is determined by the system of equations $\partial \mathcal{L}/\partial \varepsilon_F =0$ and $\partial \mathcal{L}/\partial \varepsilon_{F0} =0$ which reduces to 
%
\begin{align}
g(\varepsilon_F)(\varepsilon_F-\mu)&=0 \nonumber \\
g_0(\varepsilon_{F0})(\mu-\varepsilon_{F0})&=0,
\label{systemEF}
\end{align}
%
yielding the relation
%
\begin{align}
\varepsilon_F=\varepsilon_{F0}.
\label{EF-EF0relation}
\end{align}
%
The condition, for an extremum $\mu=\varepsilon_{F0}$ and $\mu=\varepsilon_{F}$ to be a minimum in the system with the constraint, is negativity of the determinant of the bordered Hessian
%
\begin{align}
\mathcal{H} (\varepsilon_F,\varepsilon_{F0},\mu) &\equiv \left[\begin{array}{lll}
{0} & {\partial_0\Delta N} & {\partial_1 \Delta N} \\
{\partial_0\Delta N} & {\partial_{00}\mathcal{L}} & {\partial_{0}\partial_1 \mathcal{L}} \\
{\partial_1 \Delta N} & {\partial_{1}\partial_{0}\mathcal{L}} & {\partial_{11}\mathcal{L}}
\end{array}\right] \nonumber \\
&=\left[\begin{array}{lll}
{0} & {-g_0(\varepsilon_{F0})} & {g(\varepsilon_{F})} \\
{-g_0(\varepsilon_{F0})} & {-g_0(\varepsilon_{F0})} & {0} \\
{g(\varepsilon_{F})} & {0} & {g(\varepsilon_{F})}
\end{array}\right],
\label{Hessian}
\end{align}
%
where we denoted partial derivatives $\partial_0 \equiv \partial / \partial \varepsilon_{F0}$ and $\partial_1 \equiv \partial / \partial \varepsilon_{F}$.
Namely, $\mathrm{det} (\mathcal{H}) = g_0(\varepsilon_{F0})g(\varepsilon_{F})[g(\varepsilon_{F})-g_0(\varepsilon_{F0})]<0$, taking $\varepsilon_F=\varepsilon_{F0}$, yields the condition for the minimum
%
\begin{align}
g(\varepsilon_F)<g_0(\varepsilon_F).
\label{g-g0relation}
\end{align}
%
Conditions (\ref{EF-EF0relation}) and (\ref{g-g0relation}) tell us that, in the process of band reconstruction, the Fermi energy does not change with respect to the initial one, and that the new state of the system is the minimum of energy as long as the DOS of reconstructed band is smaller than DOS of the initial band at that energy. Furthermore, from the properties of the DOS (see Fig. \ref{FigDOS}), it is evident that for dimensionality between 1 and 2, i.e. $1 \le d \le 2$, the Fermi energy $\varepsilon_F$ must be within the pseudogap between ``Lifshitz" points L$_1$ and L$_2$. On the other hand, for $d>2$, the Fermi energy may also fall in the upper band, i.e. above the ``Lifshitz" point L$_2$. However, numerical analysis shows that, as far as only the electron band contribution is concerned, the condition of new energy minimum in the reconstruction process can always be achieved.\\ 

\begin{figure}
\centerline{\includegraphics[width=\columnwidth]{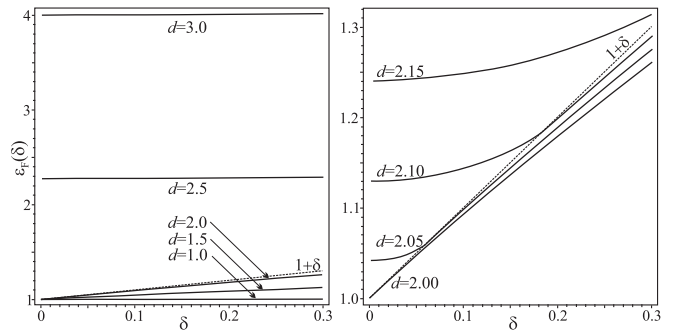}}
\caption{Dependence of the (scaled) Fermi energy on the gap parameter, $\varepsilon_F(\delta)$, obtained as a solution of Eq. (\ref{eF-equation}), for different values of dimensionality parameter $d$ (right panel presents regime around $d=2$ where qualitative change in behaviour occurs). Position of the upper "Lifshitz" point L$_2$, with (scaled) energy $1+\delta$, is depicted by the dashed line. One can see that, for $d\le 2$, position of the Fermi energy is always below L$_2$ (in the pseudogap), while for $d>2$, it can also be above it (within the upper band), depending on the value of $\delta$.} 
\label{Fig-eF}
\end{figure}

\begin{figure}
\centerline{\includegraphics[width=7.5cm]{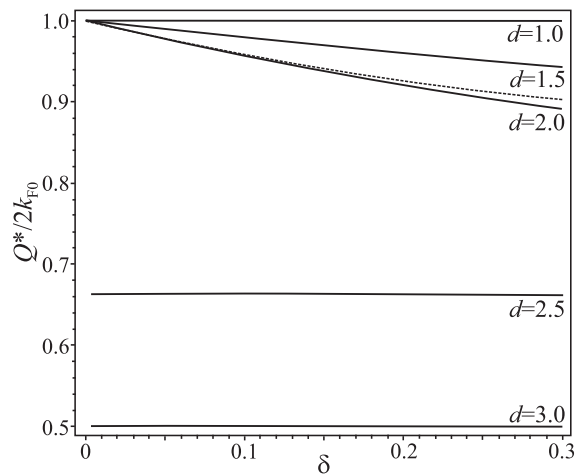}}
\caption{Optimal value of the DW wave vector (scaled by $2k_{F0}$) depending on (still free) gap parameter, $Q^*(\delta)$, for different values of dimensionality parameter $d$ obtained from Eqs. (\ref{Q-optimal}) and (\ref{eF-equation}). It is evident that for $d\le 2$, $Q^*$ develops as a $\delta-$dependent deviation from $2k_{F0}$, while, for $d>2$, it significantly differs from $2k_{F0}$, more so with increasing $d$, with practically no dependence on $\delta$. Dashed curve depicts expansion $Q^*/2k_{F0} \approx 1-\delta/2+(\delta/2)^{3/2}/\pi+3\delta^2/8$ calculated in Ref. \cite{KBR}.}
\label{Fig-Q}
\end{figure}

The initial Fermi energy is given by the system. Therefore, since $Q$ is used to scale the energy, the equation for conservation of number of electrons, $\Delta N(\varepsilon_F,\varepsilon_{F0})=0$, is in fact the equation from which we can determine optimal DW vector $Q^*$ (for a given $\delta$, at this stage treated as a parameter).
In general, the number of electrons is determined by the $d-$dimensional volume enclosed within the Fermi surface, i.e.
%
\begin{align}
N(\varepsilon)=\frac{2}{(2\pi)^d} \left( \frac{Q}{2} \right)^d \Omega_{\kappa}^{(d)}(\varepsilon).
\label{N-def}
\end{align}
%
Taking expression (\ref{volume}) for the reconstructed, and expression $\overline{\Omega}_{\kappa}^{(d)}(\varepsilon)=(\pi \varepsilon)^{d/2} / \Gamma(d/2+1)$ for non-reconstructed FS, equalizing number of electrons after and before reconstruction, $N(\varepsilon_F)=N_0(\varepsilon_F)$, leads to the equation for $\varepsilon_F$: 
%
\begin{align}
\int_{\kappa_0^{-}(\varepsilon_F)}^1 \kappa_{\rho}^{-}(\kappa_x;\varepsilon_F)^{d-1} {\mathrm d}\kappa_x +
\int_{0}^{\kappa_0^{+}(\varepsilon_F)} \kappa_{\rho}^{+} (\kappa_x;\varepsilon_F)^{d-1} {\mathrm d}\kappa_x  \nonumber \\
=\frac{\sqrt{\pi}}{2} \frac{d-1}{d} \frac{\Gamma (\frac{d-1}{2})}{\Gamma\left(\frac{d}{2}\right)} \varepsilon_F^{\frac{d}{2}}.
\label{eF-equation}
\end{align}
%
Numerically obtained solution of Eq. (\ref{eF-equation}), depicting the dependence of Fermi energy on parameter $\delta$ is shown in Fig. \ref{Fig-eF}. It is evident that for dimensionality $d \le 2$ position of the scaled Fermi energy is always below the upper ``Lifshitz" point L$_2$ (i.e. $\varepsilon_F < 1+\delta$), meaning that it is inside the pseudogap, while for $d > 2$, it may also be above it.
%
%
%
\\
Finding function $\varepsilon_F(\delta)$, as the solution of Eq. (\ref{eF-equation}), and using condition (\ref{EF-EF0relation}) as well as scaling (\ref{variables-scaling}), one directly obtains the optimal value of the DW wave vector with respect to the initial value of Fermi wave vector $k_{F0}$, depending on parameter $\delta$:
%
\begin{align}
Q^*(\delta)=\frac{2k_{F0}}{\sqrt{\varepsilon_F(\delta)}}.
\label{Q-optimal}
\end{align}
%
The $Q^*(\delta)$ dependence, for different dimensionalities $d$, is shown in Fig. \ref{Fig-Q}.
%
%
It is important to notice how the optimal DW wave vector deviates from the value $2k_{F0}$ depending on dimensionality of the system. Wave vector $Q^*=2k_{F0}$ is the value at which two initial Fermi surfaces exactly touch each other at one point. This value of $Q^*$ is exactly achieved only for $d=1$, which represents, or matches the case of perfect nesting: two Fermi surfaces, consisting only of one point, are nested to each other by unique wave vector $Q^*=2k_{F0}$ and Fermi energy is always in the middle of the gap regardless of its size. As the dimensionality of the system increases, the Fermi surfaces start to overlap (the overlap increases with increasing $d$ and depends on $\delta$ as depicted in Fig. \ref{Fig-Q}). The overlap with respect to $2k_{F0}$ is determined by factor (function) $\varepsilon_F(\delta)^{-1/2}$ in Eq. (\ref{Q-optimal}). Analytically, it can be estimated in terms of Taylor expansion, as, for example, it was done in Ref. \cite{KBR} for the 2D case ($d=2$), $\varepsilon_F(\delta)^{-1/2} \approx 1-\frac{\delta}{2}+\frac{1}{\pi}\left( \frac{\delta}{2} \right)^{3/2}$.\\

The last step in calculation of stability of the DW ground state is optimization of the order parameter $\Delta$ by minimization of Eq. (\ref{EDW-def}). 
The initial energy of the system $E_{0}(\epsilon_{F0})$ (the free-electron band with $\epsilon_{0}(\mathbf{k}) = \frac{\hbar^2 \mathbf{k}}{2m}$) is easily found (taking $\epsilon_{F0}=\epsilon_{F}$)
%
\begin{align}
E_{0}(\epsilon_{F})&= \frac{2}{(2\pi)^d} \int_{\epsilon_0(\textbf{k}) < \epsilon_{F}}  \epsilon_0(\textbf{k}) \mathrm{d}^d k \nonumber \\
&=\frac{4}{(2\pi\hbar)^d} \frac{(2\pi m)^\frac{d}{2} }{(d+2) \Gamma\left(\frac{d}{2}\right) }\epsilon_F^{\frac{d}{2}+1} \equiv E_0^{(d)}.
\label{E0}
\end{align}
%
The energy of reconstructed electron band contains contributions of lower and upper band
%
\begin{align}
E_{band}(\epsilon_F,Q,\Delta)= \frac{2}{(2\pi)^d} \left( \int_{E_-(\textbf{k}) < \epsilon_{F}}  E_-(\textbf{k}) \mathrm{d}^d k \right. \nonumber \\
\left. + \int_{E_+(\textbf{k}) < \epsilon_{F}}  E_+(\textbf{k}) \mathrm{d}^d k \right),
\label{E-rec-band-def}
\end{align}
%
which, using the scaling relations (\ref{variables-scaling}) and rotational symmetry of the FS, yields
%
\begin{align}
E_{band} & (\epsilon_F,Q,\Delta)=
\frac{4}{(2\pi)^d} \frac{\pi^{\frac{d-1}{2}}}{\Gamma \left( \frac{d-1}{2} \right)} \frac{\hbar^2}{m} \left( \frac{Q}{2} \right)^{d+2} \times \nonumber \\
& \left( \int_{\kappa_0^-(\varepsilon_F)}^{1} \mathrm{d}\kappa_x \int_0^{\kappa_{\rho}^-(\kappa_x;\varepsilon_F)} \varepsilon_-(\kappa_x,\kappa_{\rho}) \kappa_{\rho}^{d-2} \mathrm{d} \kappa_{\rho} + \right. \nonumber \\
& \left. \int_0^{\kappa_0^+(\varepsilon_F)} \mathrm{d}\kappa_x \int_0^{\kappa_{\rho}^+(\kappa_x;\varepsilon_F)} \varepsilon_+(\kappa_x,\kappa_{\rho}) \kappa_{\rho}^{d-2} \mathrm{d} \kappa_{\rho} \right).
\label{E-rec-band-1}
\end{align}
%
Integrating over $\kappa_{\rho}$ and using the relation (\ref{rho}), expressing $1+\kappa_x^2 \pm \sqrt{4\kappa_x^2+\delta^2}=\varepsilon_F-(\kappa_{\rho}^{\pm}(\kappa_x;\varepsilon_F))^2$ within expressions for $\varepsilon_{\pm}(\kappa_x,\kappa_{\rho})$ in (\ref{E-rec-band-1}), we obtain the expression
%
\begin{align}
E_{band} (\epsilon_F,Q,\Delta) &=
\frac{4}{(2\pi)^d} \frac{\pi^{\frac{d-1}{2}}}{\Gamma \left( \frac{d-1}{2} \right)} \frac{\hbar^2}{m} \left( \frac{Q}{2} \right)^{d+2} \times \nonumber \\
& \left[ \frac{\varepsilon_F}{d-1} \left( \int_{\kappa_0^-(\varepsilon_F)}^{1} \kappa_{\rho}^-(\kappa_x;\varepsilon_F)^{d-1} \mathrm{d}\kappa_x  + \right. \right. \nonumber \\
& \left. \int_0^{\kappa_0^+(\varepsilon_F)}  \kappa_{\rho}^+(\kappa_x;\varepsilon_F)^{d-1} \mathrm{d}\kappa_x \right)+ \nonumber \\
& \frac{2}{1-d^2} \left( \int_{\kappa_0^-(\varepsilon_F)}^{1} \kappa_{\rho}^-(\kappa_x;\varepsilon_F)^{d+1} \mathrm{d}\kappa_x  + \right. \nonumber \\
& \left. \left. \int_0^{\kappa_0^+(\varepsilon_F)}  \kappa_{\rho}^+(\kappa_x;\varepsilon_F)^{d+1} \mathrm{d}\kappa_x \right) \right].
\label{E-rec-band-2}
\end{align}
%
Here, recognizing the first contribution in Eq. (\ref{E-rec-band-2}) as the left-hand side of expression (\ref{eF-equation}), and subtracting the contribution of initial band energy (\ref{E0}) from the expression above, we get the change of the band energy due to the reconstruction
%
\begin{align}
\Delta E_{band} = E_0^{(d)} \left( \frac{2}{d} - \frac{2(d+2)\Gamma \left( \frac{d}{2} \right)}{\sqrt{\pi}(d+1)\Gamma \left( \frac{d+1}{2} \right)}  \frac{I_d(\varepsilon_F(\delta),\delta)}{\varepsilon_F(\delta)^{\frac{d}{2}+1}} \right),
\label{DeltE-band}
\end{align}
%
where the numeric function containing integral reads
%
\begin{align}
I_d(\varepsilon_F,\delta)= & \int_{\kappa_0^-(\varepsilon_F)}^{1} \kappa_{\rho}^-(\kappa_x;\varepsilon_F)^{d+1} \mathrm{d}\kappa_x \nonumber \\
+ & \int_0^{\kappa_0^+(\varepsilon_F)}  \kappa_{\rho}^+(\kappa_x;\varepsilon_F)^{d+1} \mathrm{d}\kappa_x.
\label{I-integral-def}
\end{align}
%
The integration in Eq. (\ref{I-integral-def}) is cumbersome to perform even numerically and it requires additional analysis. For values of $d$ close to one, when $\varepsilon_F(\delta) \rightarrow 1$, it is well approximated by
%
\begin{align}
I_d^*(\varepsilon_F=1,\delta) = \int_{0}^{1} \left( -\kappa_x^2+\sqrt{4\kappa_x^2+\delta^2} \right)^{\frac{d+1}{2}} \mathrm{d}\kappa_x.
\label{I-integral-approx}
\end{align}
%
In fact, we show that Eq. (\ref{I-integral-approx}) stays the good approximation in the whole range of dimensionality parameter $1 \le d \le 2$, i.e. as long as the Fermi energy remains in the pseudogap $1-\delta \le \varepsilon_F \le 1+\delta$ with presumably small $\delta$ (see Fig. \ref{Fig-eF}). Expanding the Eq. (\ref{I-integral-approx}) in terms of small deviation $\xi$ from $d=1$, i.e. $d=1+\xi$ expansion, as the zeroth contribution we immediately obtain $\delta^2/2 + \delta^2 \ln (4/\delta)$ contribution to the band energy characteristic for the 1D physics of Peierls transition \cite{Solyom} when the gap is fully developed (the same result would have been obtained by integrating the energy over the 1D reconstructed DOS (\ref{DOSd1})). The remaining terms with finite $\xi$, which need to be addressed numerically, are in fact corrections describing the redistribution of states due to the pseudogap formation as the dimensionality, and consequently the phase space, increases.
For dimensionality $d>2$, when also contributions of the upper band are present, we performed numerical expansion \cite{NumericalExpansion} of (rearanged) Eq. (\ref{DeltE-band}), with complete expression (\ref{I-integral-def}) taken into account, in powers of $\delta$ to the lowest important contribution
%
\begin{align}
\left( \frac{2\varepsilon_F(\delta)^2}{d} - \frac{2(d+2)\Gamma \left( \frac{d}{2} \right)}{\sqrt{\pi}(d+1)\Gamma \left( \frac{d+1}{2} \right)}  \frac{I_d(\varepsilon_F(\delta),\delta)}{\varepsilon_F(\delta)^{\frac{d}{2}-1}} \right) \approx \nonumber \\
 -\frac{1}{\lambda_c^{(d)}}\delta^2 + \mathcal{C}^{(d)}\delta^3,
\label{Expansion}
\end{align}
%
where numerically obtained constants $\lambda_c^{(d)}$ and $\mathcal{C}^{(d)}$ depend on $d$. Their numerical values determine the quantitative aspects of the problem, while for the qualitative aspect, i.e. the existence of the DW transition, it is important that they are positive. Here one can directly see the negative energy contribution (the first term at the right-hand side) which may stabilize the new ground state in the reconstruction process. The results of this expansion overlap with the results of approximation (\ref{I-integral-approx}) in the interval $1.8<d \le 2$, so using both of them we can cover the whole $d \in [1,3]$ range.\\

So far $\Delta$ was treated as a free parameter, thus depicting the dependence of all quantities on $\delta$, regardless of its scaling, was convenient. However, in the final step, we perform an optimization with respect to $\Delta$ and we choose some fixed energy scale, e.g. the Fermi energy $\epsilon_F$. The ``elastic" energy term from Eq. (\ref{EDW-def}) we express as  
%
\begin{align}
\frac{\Delta^2}{G} \equiv E_0^{(d)} \frac{\left( \Delta / \epsilon_F \right)^2}{\lambda^{(d)}},
\label{G-scaling}
\end{align}
%
where 
%
\begin{align}
\lambda^{(d)} \equiv E_0^{(d)} \frac{G}{\epsilon_F^2}
\label{Lambda}
\end{align}
%
is the dimensionless coupling constant (taking the role of $G$) which incorporates dimensionality $d$.\ 

\begin{figure}
\centerline{\includegraphics[width=\columnwidth]{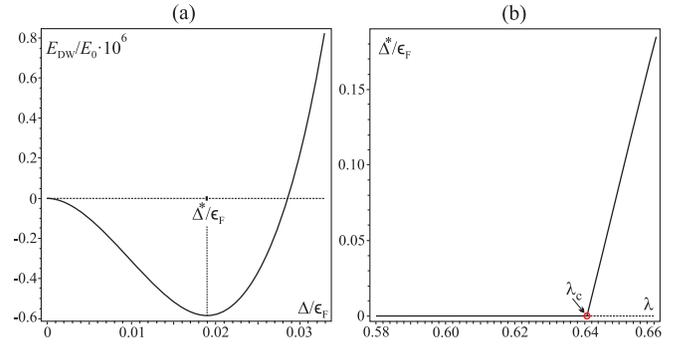}}
\caption{(a) The DW condensation energy $E_{DW}(\Delta)$, given by Eq. (\ref{EDW-final}), exhibits minimum at optimal value of the order parameter $\Delta^*$ if the coupling constant is larger than critical for a given dimensionality. Here, for illustration, we chose $d=2.1$ and $\lambda=0.642$, with $\lambda_c \approx 0.64$ provided. (b) The optimal value of the order parameter depending on the coupling constant, $\Delta^*(\lambda)$. We marked the bifurcation (critical) point at $\lambda=\lambda_c$ separating two stable types of solutions: $\Delta^*=0$ for $\lambda\le\lambda_c$, and $\Delta^*\ne 0$ for $\lambda\ > \lambda_c$. Here, for illustration, we chose $d=2.1$ yielding $\lambda_c \approx 0.64$.}
\label{Fig-DeltaMinimum}
\end{figure}

It is easiest to demonstrate the onset of the DW transition first using the approximation (\ref{Expansion}) (naturally, within the range of its validity), which provides rather good phenomenological description. Using relation $\delta/\varepsilon_F = \Delta / \epsilon_F$, in which we cancel the old scaling, we can write expression (\ref{EDW-def}) for the DW condensation energy
%
\begin{align}
\frac{E_{DW}}{E_0^{(d)}} = \left( \frac{1}{\lambda^{(d)}} - \frac{1}{\lambda_c^{(d)}} \right) \left( \frac{\Delta}{\epsilon_F} \right)^2 + \mathcal{C}^{(d)} \varepsilon_F^{(d)} \left( \frac{\Delta}{\epsilon_F} \right)^3.
\label{EDW-final}
\end{align}
%
Here, due to conversion $2m\Delta/(\hbar^2 Q/2)^2=\varepsilon_F\Delta/\epsilon_F$, appears a factor $\varepsilon_F^{(d)}=\varepsilon_F(\delta \rightarrow 0)$  which has to be obtained numerically, with $\delta-$dependence neglected (for $d>2$ and presumably small $\delta$ it is very weak and, also, any $\delta-$dependence leads to higher order corrections than considered), but important dependence on $d$ (see Fig. \ref{Fig-eF}).
Based on expansion (\ref{Expansion}) we get physical insight in the nature of the DW ground state: the minimum of function (\ref{EDW-final}) exists, i.e., it is possible to stabilize the DW state, if the coupling constant $\lambda$ is greater than critical, i.e., $\lambda > \lambda_c$ for a given dimensionality $d$ (see Fig. \ref{Fig-DeltaMinimum}(a)).
With condensation energy expressed in terms of expansion, it is easy to find optimal order parameter $\Delta^*$ simply by finding the zero-point of derivative of Eq. (\ref{EDW-final})
%
\begin{align}
\frac{\Delta^*}{\epsilon_F} \approx
\begin{cases} 
0 & \lambda^{(d)} \le \lambda_c^{(d)} \\
\frac{2}{3\mathcal{C}^{(d)} \varepsilon_F^{(d)}} \left( \frac{1}{\lambda_c^{(d)}} - \frac{1}{\lambda^{(d)}} \right) & \lambda^{(d)} > \lambda_c^{(d)}.
\end{cases}
\label{Delta-optimal}
\end{align}
%
This solution features a typical bifurcation behaviour with change of stability: for $\lambda \le \lambda_c$, the solution with the zero order parameter ($\Delta^*=0$) is stable while, for $\lambda > \lambda_c$, the solution with the zero order parameter loses stability, and the solution with a finite order parameter ($\Delta^* \ne 0$) is stabilized (see Fig. \ref{Fig-DeltaMinimum}(b)).\\
%
%
The more detailed microscopic understanding of the transition process and critical coupling is provided using the approximation (\ref{I-integral-approx}) (again, within its range of validity) in Eqs. (\ref{DeltE-band}) and (\ref{EDW-def}) with the afore-mentioned change $\delta/\varepsilon_F = \Delta / \epsilon_F$. Taking the derivative of the DW energy with respect to $\Delta$ and equalizing it with zero leads to the equation for the optimal order parameter
%
\begin{align}
\frac{\Delta}{\epsilon_F} \left(\frac{(d+2)\Gamma \left( \frac{d}{2} \right)}{2\sqrt{\pi}\Gamma \left( \frac{d+1}{2} \right)}  \int_0^1 {\frac{\left( -\kappa_x^2+\sqrt{4\kappa_x^2+\left(\frac{\Delta}{\epsilon_F}\right)^2} \right)^{\frac{d+1}{2}}}{\sqrt{4\kappa_x^2+\left(\frac{\Delta}{\epsilon_F}\right)^2}} \mathrm{d}\kappa_x } \right. \nonumber \\
\left. - \frac{1}{\lambda^{(d)}} \right) =0.
\label{DeltaEquation}
\end{align}
%
Here, within the $1<d<2$ interval, the conversion factor appearing near $\Delta/\epsilon_F$ is approximately $\varepsilon_F \approx 1$ for small $\delta$ ($\delta-$dependence leads again only to higher-order corrections than considered), thus it is omitted from the equation for simplicity (see Fig. \ref{Fig-eF}).
As mentioned above, one stable solution is $\Delta=0$, originating from the first term in Eq. (\ref{DeltaEquation}) for $\lambda \le \lambda_c$, while the zero of the second term determines finite $\Delta \ne 0$ for a given $\lambda \ge \lambda_c$. Furthermore, the zero of the second term determines the value of critical coupling: by setting $\lambda = \lambda_c \Leftrightarrow \Delta =0$, it immediately yields
%
\begin{align}
\lambda_c^{(d)} =\frac{4\sqrt{\pi}\Gamma \left( \frac{d+1}{2} \right)}{(d+2)\Gamma \left( \frac{d}{2} \right)}
 \left(\int_0^1 {\frac{\left( 2\kappa_x-\kappa_x^2 \right)^{\frac{d-1}{2}}}{\kappa_x} \mathrm{d}\kappa_x } \right)^{-1}.
\label{LambdaCrit}
\end{align}
%
The limiting cases of validity interval are easily obtained: for example in the 1D case ($d=1$) we get $\lambda_c^{(1)}=\frac{4}{3}\left(\int_0^1 \frac{\mathrm{d}\kappa_x}{\kappa_x} \right)^{-1} \rightarrow 0$ as expected \cite{Solyom}, while in the 2D case ($d=2$), directly evaluating the integral, we get $\lambda_c^{(2)}= \left( 1+\frac{2}{\pi} \right)^{-1}$ as obtained in Ref. \cite{KBR}.\

\begin{figure}
\centerline{\includegraphics[width=7.0cm]{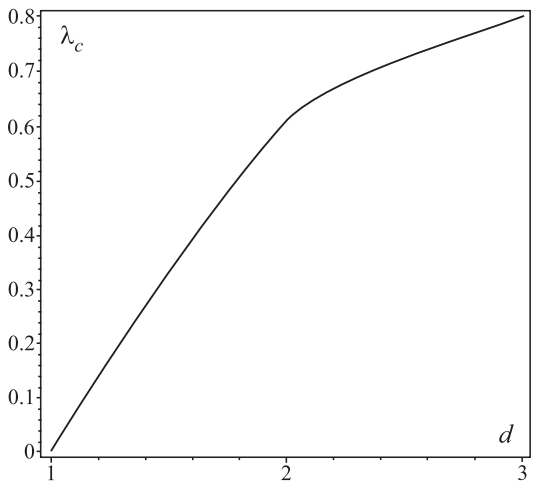}}
\caption{The critical coupling strength $\lambda_c^{(d)}$ of the DW ordering depending on the dimensionality of the system $d$.}
\label{Fig-LambdaC}
\end{figure}

As we can see, the critical coupling strength $\lambda_c^{(d)}$ depends strongly on the dimensionality of the system. The result of numerical analysis, based both on approximation (\ref{I-integral-approx}) and expansion (\ref{Expansion}), leading to expressions (\ref{LambdaCrit}) and (\ref{EDW-final}) respectively in their own ranges of validity, is presented in Fig. \ref{Fig-LambdaC}.\
%
%

The origin of such a behaviour lays in the structure of electron DOS accompanying the band reconstruction process. In the strictly 1D case ($d=1$), the gap is real (electron DOS exactly vanishes between ``Lifshitz" points L$_1$ and L$_2$), as it is already mentioned in the Introduction and known from the literature on Peierls transition \cite{Solyom}. The presence of the gap leads to the characteristic contribution $E_{band} \sim \Delta^2 \ln{(\epsilon_F/\Delta)}$ in the band energy, which in turn gives optimal order parameter $\Delta^* \sim \epsilon_F \exp{(-1/\lambda)}$. Therefore, even for arbitrarily small coupling, the DW ground state is stable ($\Delta^* \ne 0$) in the 1D case. As the dimensionality is increased ($d>1$), the spectrum between the ``Lifshitz" points is not fully gapped any more and the pseudogap arises instead, i.e. the number of states decreases compared to the initial state, but it does not vanish exactly. The consequence is an absence of the DW transition in the zero-coupling limit. However, the lower the dimensionality is, the more states are redistributed to the lower energy around L$_1$, thus smaller, but finite coupling constant is required to stabilize the DW ground state. With increasing dimensionality, states within the pseudogap are distributed more and more at higher energies which consequently increases $\lambda_c^{(d)}$ required to stabilize the DW.

\bigskip
\textbf{V. Conclusions}
\bigskip

We have studied stability of an electron system with closed FSs with respect to the spontaneous formation of an uniaxial DW, in dimensional crossover between 1D and 3D systems at zero temperature. In a way, one may say that we generalize the aspects of Peierls transition to a higher-than-one dimension. We assume the DW with such periodic modulation that causes topological reconstruction of the initial FS, from the set of $d-$dimensional (hyper)spheres with radius $k_F$ in the reciprocal space, to the FS with open topology. In order to achieve such topology, the DW wave vector $Q$ should be close to $2k_F$. In this process, two bands are formed in the electron energy spectrum, lower with hyperbolic (L$_1$) and upper  with elliptical (L$_2$) point for $d>1$. These points, at energies $E_{L1}$ and $E_{L2}$ respectively, are peculiar (``Lifshitz") points distinguished in energy by the gap parameter $2\Delta$ (which appears as the order parameter of the DW state). The influence of the reconstruction process and opening of the gap in the electron spectrum is tracked through the calculation of the electron DOS (Fig. \ref{FigDOS}). The $d=1$ is the well known Peierls case with full gap opened between $E_{L1}$ and $E_{L2}$. The system is unstable with respect to formation of the DW ground state down to the zero-coupling limit in terms of interactions. However, as the dimensionality increases, i.e. $d>1$, instead of gap, the pseudogap with depleted, but still present states between $E_{L1}$ and $E_{L2}$ is formed, with less and less states redistributed to lower energies as $d$ increases. Calculations of the DW condensation energy show that the DW ground state is stable if the interaction coupling constant is larger than the critical value, which depends on dimensionality of the system (Fig. \ref{Fig-DeltaMinimum}), i.e. the transition to the DW state manifests itself as a kind of quantum phase transition in the coupling strength space. The critical coupling constant monotonously increases from the zero value in the 1D case to higher values for higher dimensionality $d$ (Fig. \ref{Fig-LambdaC}). Consequently, the higher the dimensionality of the system is, the more ``difficult" it is for electrons to establish the DW ground state, i.e. the stronger interactions are required. The DW wave vector also changes with dimensionality. From the strict $2k_F$ value in the 1D case, which relates FSs into strict touching, with increasing dimensionality the FSs start to overlap more (Fig. \ref{Fig-Q}). This overlap, determined by the optimal DW wave vector, appears in the nonmonotonous way: For $1 < d \le 2$, $Q=2k_F(1-corr^{(d)}(\Delta))$, i.e. the overlap of the FSs is determined by presumably small $\Delta-$dependent correction to touching,  which depends also on $d$. For $2 < d \le 3$, $Q$ changes significantly with $d$, while $\Delta-$dependence is practically negligible, ending at $Q \approx k_F$ for $d=3$ which provides significant optimal overlap, comparable to the size of the FS. From the aspect of condensation energy, the larger the overlap, proportionally smaller the condensation energy of the DW is. Clearly, in the competition between the Peierls scenario and Lifshitz scenario, the first-mentioned wins as long as the system provides necessary assets, i.e. the FSs with the property of nesting. However, here we show that transition into the DW ground state is possible in the complementary ("antinesting") limit and address the assets that the system needs - the critical coupling strength.\\
The present analysis supplements the one in the Ref. \cite{KBR}, done for the 2D case, in our intention to give an initial framework for understanding the nature of CDWs experimentally observed in a number of materials, like high$-T_c$ cuprates or graphite intercalates. Both of them are materials with closed, rather isotropic FSs in a plane, far from the nesting regime. On the other hand, they are essentially 3D materials, with highly pronounced anisotropy perpendicular to the mentioned plane, here modelled by the noninteger parameter $d$. The forthcoming step from this general description would be introducing the finite temperature as well as modeling the particular material dispersions and the DW geometries, as well as inserting the real material parameters into the model to fit the phase diagram. Also, addressing the presumably important effects of commensurability of the new DW ordering with the underlying lattice, which are outside of the scope of this work, might be important.  

\bigskip

\textbf{\emph{Acknowledgement}}. This work was supported by the Croatian Science Foundation, project IP-2016-06-2289,
and by the QuantiXLie Centre of Excellence, a project cofinanced by the Croatian Government and European Union
through the European Regional Development Fund - the Competitiveness and Cohesion Operational Programme
(Grant KK.01.1.1.01.0004).\\

*e-mail: dradic@phy.hr


\begin{thebibliography}{99}
\bibitem{123DEG} Inserting $d=1$, 2, 3 into Eq. (\ref{DOS-g0}), we get $g_0(E)=\frac{\sqrt{2m}}{\pi \hbar} \frac{1}{\sqrt{E}}$, $g_0(E)=\frac{m}{\pi \hbar^2}$, $g_0(E)=\frac{m\sqrt{2m}}{\pi^2 \hbar^3} \sqrt{E}$ for 1D, 2D and 3D case respectively.
\bibitem{derivative} Derivative with respect to energy $\varepsilon$ must also take into account limits of integration in expression (\ref{volume}), i.e.
$\frac{\partial}{\partial \varepsilon} \int_{\alpha(\varepsilon)}^{\beta(\varepsilon)} f(\kappa, \varepsilon) \mathrm{d} \kappa=\beta^{\prime}(\varepsilon) f[\beta(\varepsilon), \varepsilon]-\alpha^{\prime}(\varepsilon) f[\alpha(\varepsilon), \varepsilon]+\int_{\alpha(\varepsilon)}^{\beta(\varepsilon)} \frac{\partial f(\kappa, \varepsilon)}{\partial \varepsilon} \mathrm{d} \kappa$. For $d=1$ the only contribution to the DOS (\ref{DOSd1}) comes from first two terms (limits of integration), while for $d>1$ the only contribution in (\ref{DOSdgt1}) comes from the third term (integrand).
\bibitem{NumericalExpansion} We used function "find-fit()" in the program package Sage with accuracy of $10^{-5}$ in mean squared deviation.
\end{thebibliography}
\end {document}